\documentclass[aps,prl,twocolumn,superscriptaddress]{revtex4-1}

\usepackage{amsmath}
\usepackage{amssymb}
\usepackage{wasysym}
\usepackage{graphicx}
\usepackage{hyperref}
\usepackage{color}
\usepackage{physics}
\usepackage{siunitx}
\usepackage[english,nomargin,inline,marginclue,draft]{fixme}
\fxusetheme{colorsig}
\FXRegisterAuthor{cg}{acg}{CG}  % now one can use \cgnote \cgwarning \cgerror
\FXRegisterAuthor{tr}{atr}{TR}
\makeatletter
\renewcommand*\FXLayoutInline[3]{%
  {\@fxuseface{inline}\ignorespaces{\color{fx#1}[#3: #2]}}}
\makeatother
\makeatletter
\renewcommand*{\@fnsymbol}[1]{\ensuremath{\ifcase#1\or *\or *,\dagger\or  \ddagger\or
    \mathsection\or \mathparagraph\or \|\or **\or \dagger\dagger
    \or \ddagger\ddagger \else\@ctrerr\fi}}
\makeatother

\begin{document}

\title{Many-body delocalization  in the \\ presence of a quantum bath}

\author{Antonio Rubio-Abadal}
\thanks{These authors contributed equally to this work}
\affiliation{Max-Planck-Institut f\"{u}r Quantenoptik, 85748 Garching, Germany}

\author{Jae-yoon Choi}
\email{jaeyoon.choi@kaist.ac.kr}
\affiliation{Max-Planck-Institut f\"{u}r Quantenoptik, 85748 Garching, Germany}
\affiliation{Department of Physics, Korea Advanced Institute of Science and Technology, Daejeon 3414, Korea}

\author{Johannes Zeiher}
\affiliation{Max-Planck-Institut f\"{u}r Quantenoptik, 85748 Garching, Germany}

\author{Simon Hollerith}
\affiliation{Max-Planck-Institut f\"{u}r Quantenoptik, 85748 Garching, Germany}

\author{Jun Rui}
\affiliation{Max-Planck-Institut f\"{u}r Quantenoptik, 85748 Garching, Germany}

\author{Immanuel Bloch}%
\affiliation{Max-Planck-Institut f\"{u}r Quantenoptik, 85748 Garching, Germany}
\affiliation{Fakult\"{a}t f\"{u}r Physik, Ludwig-Maximilians-Universit\"{a}t M\"{u}nchen, 80799 M\"{u}nchen, Germany}%

\author{Christian Gross}%
\affiliation{Max-Planck-Institut f\"{u}r Quantenoptik, 85748 Garching, Germany}

\date{\today}

%%%%%%%%%%%%%%%%%%%%%%%%%%%%%%%%%%%%%%%%%%%%%%%%%%%%%%%%%%%%%%%%%%%%%%%
%                        Summary paragraph                            %
%%%%%%%%%%%%%%%%%%%%%%%%%%%%%%%%%%%%%%%%%%%%%%%%%%%%%%%%%%%%%%%%%%%%%%%

\begin{abstract} 

Closed generic quantum many-body systems may fail to
thermalize under certain conditions even after long times, a phenomenon called many-body localization (MBL). Numerous studies support the stability of the MBL phase in strongly disordered one-dimensional systems. However, the situation is much less clear when a small part of the system is ergodic, a scenario which also
has important implications for the existence of many-body localization in
higher dimensions. Here we address this question experimentally using a
large-scale quantum simulator of ultracold bosons in a two-dimensional optical
lattice. We prepare two-component mixtures of varying relative population and
implement a disorder potential which is only experienced by one of the
components. The second non-disordered ``clean'' component plays the role of a bath of adjustable size that is collisionally coupled to the
``dirty'' component. Our experiments show how the dynamics of the dirty component, which, when on its own, show strong evidence of localization, become affected by the coupling to the clean component. For a high clean population, the clean component appears to behave as an effective bath for the system which leads to its delocalization, while for a smaller clean population, the ability of the bath to destabilize the system becomes strongly reduced. Our results reveal how a finite-sized quantum system can bring another one towards thermalization, in a regime of complex interplay between disorder, tunneling and intercomponent interactions. They provide a new benchmark for effective theories aiming to capture the complex physics of MBL in the weakly localized regime.

\end{abstract}

\maketitle

Typical quantum many-body systems evolve into a locally thermal state after
driven out of equilibrium by a global quench~\cite{rigol2008}. This quantum
version of thermalization is explained by the eigenstate thermalization
hypothesis, which postulates that small subsystems are described by a thermal
density matrix even for individual many-body eigenstates of the global
system~\cite{deutsch1991, srednicki1994, tasaki1998, kaufman2016}. Quantum
thermalization can, however, fail generically in systems exhibiting quenched
disorder~\cite{anderson1958, basko2006,gornyi2005} when the strength of the
disorder is large enough to prevent efficient spreading of
entanglement~\cite{nandkishore2015}. Non-thermalizing behaviour and strong
indication for the existence of a many-body-localized phase have been
observed experimentally in several systems, in one~\cite{schreiber2015,
smith2016, roushan2017, wei2018, xu2018} as well as in two
dimensions~\cite{choi2016, bordia2017}. An efficient delocalization of such MBL systems can be induced by coupling them to an external dissipative bath, which will ultimately restore thermalization. But similar processes can take place even in perfectly isolated systems, when a large enough part of the system is ergodic. Specifically, the many-body eigenstates in some parts of the energy spectrum may obey the eigenstate thermalization hypothesis while in other parts remain localized~\cite{li2015a,deroeck2016,lueschen2017} and spatial rare regions of low disorder may form local ergodic inclusions~\cite{agarwal2017, deroeck2017} that can trigger a destabilizing avalanche, a common argument against the existence of an MBL phase in higher dimensions. An understanding of the mechanisms which lead to thermalization and to the destruction of MBL is thus of central importance to verify the robustness of the MBL phase. Such processes pose a challenge for both numerical and experimental methods, since their identification require the study of long evolution times in large quantum systems. 

\begin{figure*}
\centering
\includegraphics[width=150 mm]{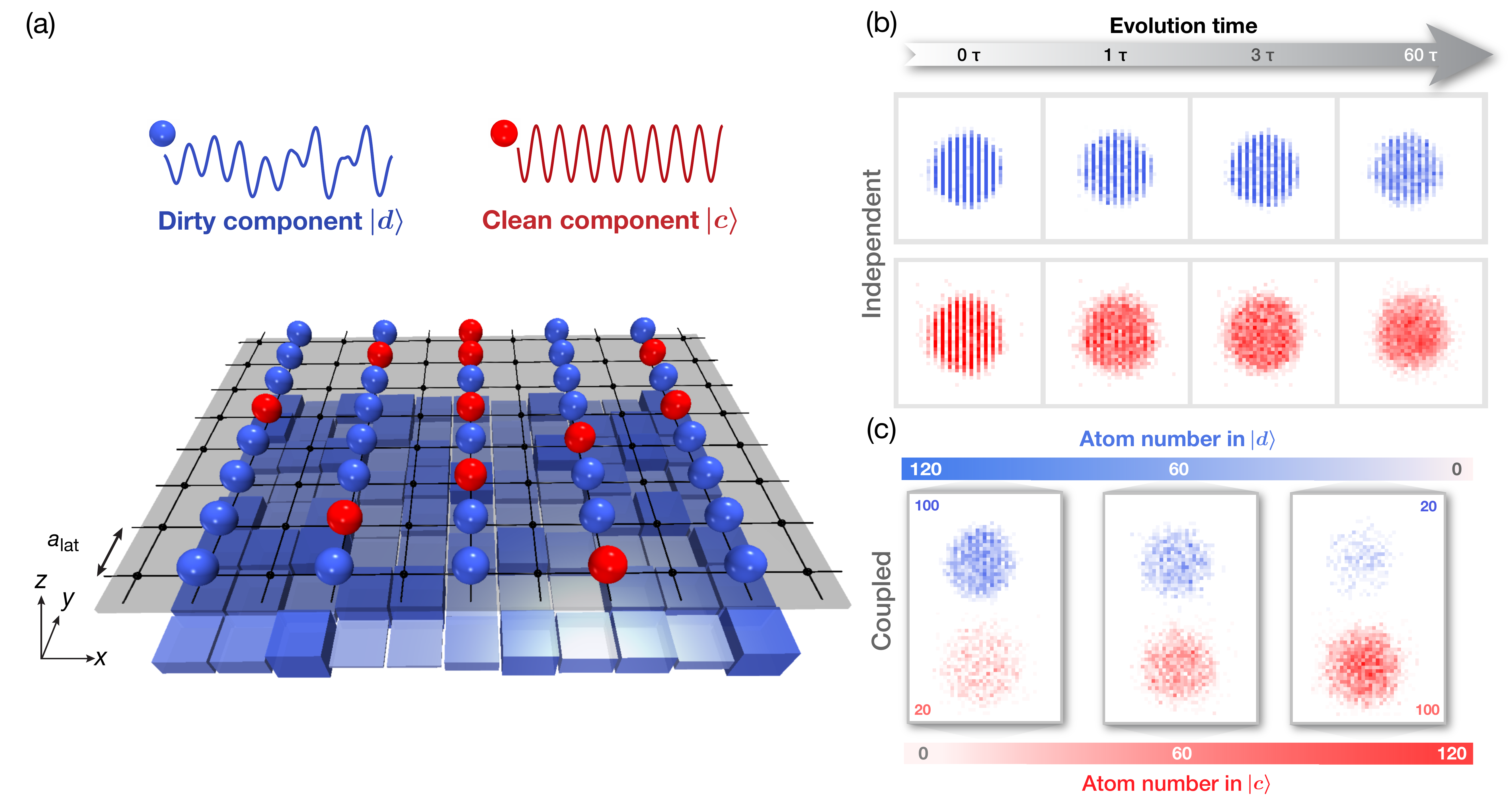}
\caption{ \label{fig:1}
\textbf{Schematic description of the experiment.} (a) Illustration of the two-component mixture at the beginning of the experimental sequence. The system consists of a square optical
lattice with species-dependent disorder in which two different bosonic
hyperfine spin states are prepared. The clean component (red, $\ket{c}$)
experiences only the lattice potential, while the dirty component (blue, $\ket{d}$) is
additionally affected by a random on-site potential (blue boxes of different lightness). The initial state is prepared with a short-scale density modulation along the \textit{x}-direction. (b) Dynamics of the mean density distribution for systems prepared independently with only either one of the two components.  The dirty component partially preserves the initial-state density modulation for a sufficiently high
disorder strength (in blue, for $\mathrm{\Delta} = 28\,J$),  which is a signature of the breakdown of ergodicity and the formation of an MBL phase. The clean component
$\ket{c}$, in contrast, relaxes in a few tunneling times to a state with no clear density modulation (in red). (c) Mean state-selective density distribution of a mixture after a long evolution time.
The measurements were taken after $281\,\tau$ evolution for three exemplary component fractions.}
\end{figure*}

An experiment which allows to probe such delocalizing phenomena in a highly controlled setting, and in the complex regime near the transition between the ergodic and localized phases, consists of an interacting two-component mixture composed of a ``dirty'' component in a random potential, and a ``clean'' component insensitive to the disorder~\cite{nandkishore2015a, hyatt2017, marino2018}. In such a hybrid system, the clean component, which on its own would quantum thermalize, can be viewed as a quantum bath with a tunable number of degrees of freedom~\cite{huse2015, banerjee2016, bordia2016}. In particular the small-bath regime can be realized, a very different scenario compared to the theoretically~\cite{nandkishore2014,johri2015,levi2016a,fischer2016} and experimentally studied~\cite{lueschen2017a} coupling of an MBL system to a classical bath at infinite coupling bandwidth. Under some conditions, a small quantum bath might even fail to thermalize the whole system. Such a breakdown of thermalization can happen as a consequence of the bath becoming localized via the intercomponent interactions, which can play the role of an effective disorder (an MBL proximity effect~\cite{nandkishore2015a, hyatt2017}). In this work, we use a quantum-gas microscope to prepare an out-of-equilibrium state in a disordered potential, and to measure its dynamics for long time scales beyond 1000 tunneling times. By introducing a second species insensitive to the disorder, we realize the setting described above. While the dirty component shows strong indication of localization in the absence of a bath, the introduction of a large enough number of clean atoms alters the dynamics qualitatively, and the signs of localization vanish eventually.

Our experimental system consists of a square optical lattice with lattice spacing $a_{\mathrm{lat}}=\SI{532}{nm}$ in which we load an ultracold cloud of $^{87}\mathrm{Rb}$ atoms. By preparing the atoms in two different hyperfine states, the dirty $\ket{d}=\ket{F=2, m_{F}=-2}$ and clean $\ket{c}=\ket{F=1, m_{F}=-1}$, a two-species bosonic mixture is obtained, which features almost equal
inter- and intraspecies interactions $U_{dc} \simeq U_{cc} \simeq U_{dd} \equiv U$
\cite{fukuhara_microscopic_2013}. We optically induce a state-dependent on-site disorder potential
$\delta_\textit{\textbf{i}}$, which only affects the dirty
$\ket{d}$ component, thereby breaking SU(2) symmetry (see Supplementary
Information~\cite{SI}). This system can be described by a two-species disordered Bose-Hubbard Hamiltonian:
\begin{align}
\hat{H} =
&-J  \sum_{\langle  \textit{\textbf{i}}, \textit{\textbf{j}} \, \rangle, \sigma} \hat{a}_{\textit{\textbf{i}},\sigma}^{\dagger} \, \hat{a}_{\textit{\textbf{j}}, \sigma}
+ \frac{U}{2} \sum_{\textit{\textbf{i}}, \sigma} \hat{n}_{\textit{\textbf{i}},\sigma}(\hat{n}_{\textit{\textbf{i}},\sigma}-1) \\
&+  U\sum_{\textit{\textbf{i}}} \hat{n}_{\textit{\textbf{i}},d}\,\hat{n}_{\textit{\textbf{i}},c} 
+  \sum_{\textit{\textbf{i}}, \sigma} V_\textit{\textbf{i}} \,\hat{n}_{\textit{\textbf{i}}, \sigma}
+ \sum_{\textit{\textbf{i}}} \delta_\textit{\textbf{i}}\, \hat{n}_{\textit{\textbf{i}},d}, \nonumber
\end{align}
where $\hat{a}_{\textit{\textbf{i}},\sigma} $, $\hat{a}_{\textit{\textbf{i}},\sigma}^{\dagger}$ and $\hat{n}_{\textit{\textbf{i}},\sigma} $ denote the annihilation, creation and number operators for a particle in state $\sigma \in \{ c,d \}$ at site $\textit{\textbf{i}}$ of our 2D system [$\textit{\textbf{i}}=(i_{x},i_{y})$]. The first sum indicates the hopping between nearest-neighbour sites $\langle \textit{\textbf{i}},\textit{\textbf{j}} \,\rangle$ with a state-independent tunneling amplitude $J$, and $V_{\textit{\textbf{i}}}$ characterizes the harmonic trapping potential.

To prepare the out-of-equilibrium initial state for the experiments in this work, we start with a unity-filling Mott
insulator in the atomic limit and remove the atoms on every second column
such that $N=\num{124(12)}$ atoms remain (see Fig.~\ref{fig:1}(b)).  We then prepare a fraction of the atoms in state $\ket{c}$ via a
microwave pulse before imposing the disorder potential and quenching the
optical-lattice tunneling down to $J/\hbar=2\pi \times \SI{24.8}{Hz}$ and the interactions
to $U=24.4\,J$ (see Fig.~\ref{fig:1} and
Supplementary Information for details~\cite{SI}). The random disorder potential is chosen to be different for each
experimental realization and its distribution is approximately Gaussian with a full
width at half maximum of $\Delta=28\,J$~\cite{choi2016}. After the lattice quench, we allow the system to evolve for
up to $t\simeq 1100\, \tau$, where $\tau= \hbar /J$ is the characteristic tunneling
time. After the evolution, the lattices are increased to their maximum depth, in order
to freeze the spatial distribution and image the atomic
occupations on each individual lattice site~\cite{sherson2010}. For the chosen value of the disorder strength $\Delta$, the density distribution of the dirty component eventually reaches a steady state, retaining a signature of the initial density modulation, while the clean one quickly becomes featureless (see Fig.~\ref{fig:1}(b)). The access to the individual site occupations allows us to characterize these dynamics by 
tracking the imbalance $\mathcal{I} =
(N_{\mathrm{e}}-N_{\mathrm{o}})/(N_{\mathrm{e}}+N_{\mathrm{o}})$, where
$N_{\mathrm{e},\mathrm{o}}$ are the occupation numbers on sites at even and
odd columns. 
The use of the short-distance density modulation efficiently probes localization at short length scales.

\begin{figure}
\centering \includegraphics{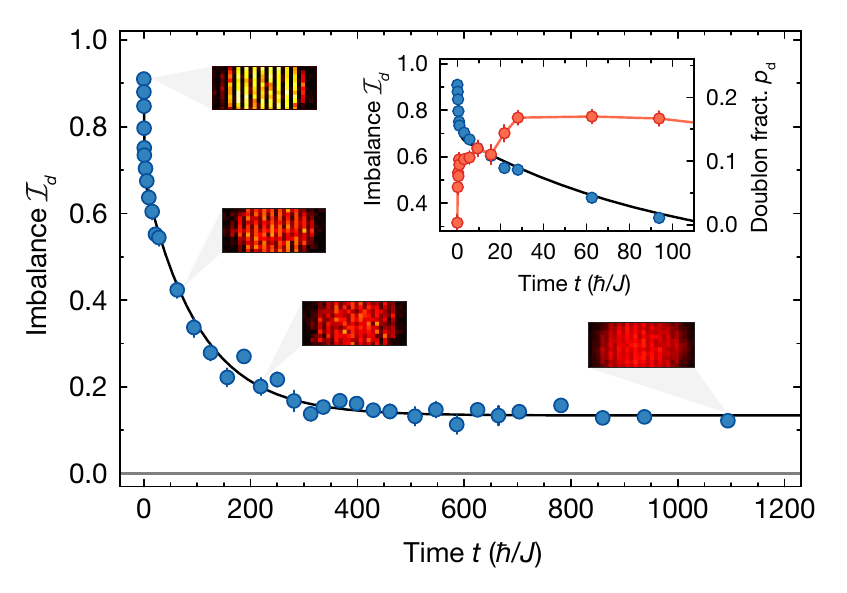} \caption{ \label{fig:2}
\textbf{Dynamics of the dirty component alone.} Evolution of the imbalance
$\mathcal{I}_d$ (points) for a system composed only of the dirty component $|d\rangle$. The blue points show the measured data and the solid line a fit of the points to the sum of two exponentials including an offset. The imbalance decreases monotonically, albeit on distinct time scales. An initial decay, where interactions are negligible, is followed by a slower second time scale
governed by doublon relaxation. 
After $\sim 300\,\tau $ a quasi-steady state of
finite imbalance $\mathcal{I}_d\approx0.13$ is reached. The four rectangular boxes show
the averaged density distribution in the trap center (black-red-yellow color scale)
at times $0\,\tau$, $63\,\tau$, $219\,\tau$ and $1094\,\tau$, displaying the
reduction in the imbalance. In the inset, we show the short-time dynamics of
the imbalance $\mathcal{I}_d$ (blue) and the doublon population
$p_{\mathrm{d}}$  (red), defined as the fraction of atoms in doubly-occupied
lattice sites. Notably, the doublon formation rate changes markedly between
the two dynamical regimes of the imbalance decay. The error bars represent one
standard deviation of the mean. }
\end{figure}

We start by preparing a state in which all particles are in the dirty component, i.e. in the absence of the atomic bath, and  measure the evolution of its imbalance $\mathcal{I}_{d}$. Over a few hundreds of tunneling times, we observe a decrease of the initial imbalance from $\mathcal{I}_{d}=0.91(1)$, to a long-time steady value of $\mathcal{I}_{d}\approx 0.13$ -- a signature of MBL. The measured dynamics can be phenomenologically well described by the sum of two exponentials with vastly different time constants and a stationary offset. During the first identified period, characterized by a decay time of $0.6(1)\,\tau$, the atoms expand freely into empty sites. In the following period of much slower dynamics (decay time $103(6)\,\tau$) interactions are important and, together with the even longer time dynamics, this time scale constitutes the focus of our analysis. In addition, we resolve the formation of doubly occupied sites (doublons) during the relaxation dynamics, starting from an initially doublon-free case, which exhibits a rapid growth and subsequent saturation after the quench. This effect requires both interactions and disorder, since in a disorder-free lattice with the same parameters such a dynamical doublon formation is strongly suppressed by the interactions. The qualitative behavior of these dynamics is reproduced by small-system numerical simulations (see Supplementary Information~\cite{SI}). Furthermore, recent numerical work on the 2D disordered Bose-Hubbard model has found signatures of MBL by studying individual eigenstate properties at parameters consistent with this work~\cite{Wahl_MBL}. Nonetheless, the finite coupling to the environment in our system, inevitable in any experiment, becomes increasingly important at longer times. Its dominant effect is an atom loss of $15\%$ after $600\,\tau$. The effect of such loss on localization is not entirely clear, but in our experiments it does not seem to cause strong delocalization. No total relaxation of the imbalance has been observed even for the longest measured times, which are among the longest times probed in any MBL-related experiment so far. The data of Fig. 2 show that the time scale of any potential subsequent relaxation is large and well separated from the characterized initial decay. In addition, an exponential fit of the data after $500\,\tau$, combined with a bootstrap analysis, allows us to bound any further relaxation to be $t_{3}>2300\,\tau$ with $92\%$ confidence. This lower bound is our sensitivity limit for the relaxation, in the sense that any slower decay process cannot be distinguished from true localization. The large separation of timescales between the initial decay and any potential further decay is, however, large, rendering the observed plateau a useful indication for the presence of MBL in the system.

\begin{figure*}
\centering
\includegraphics{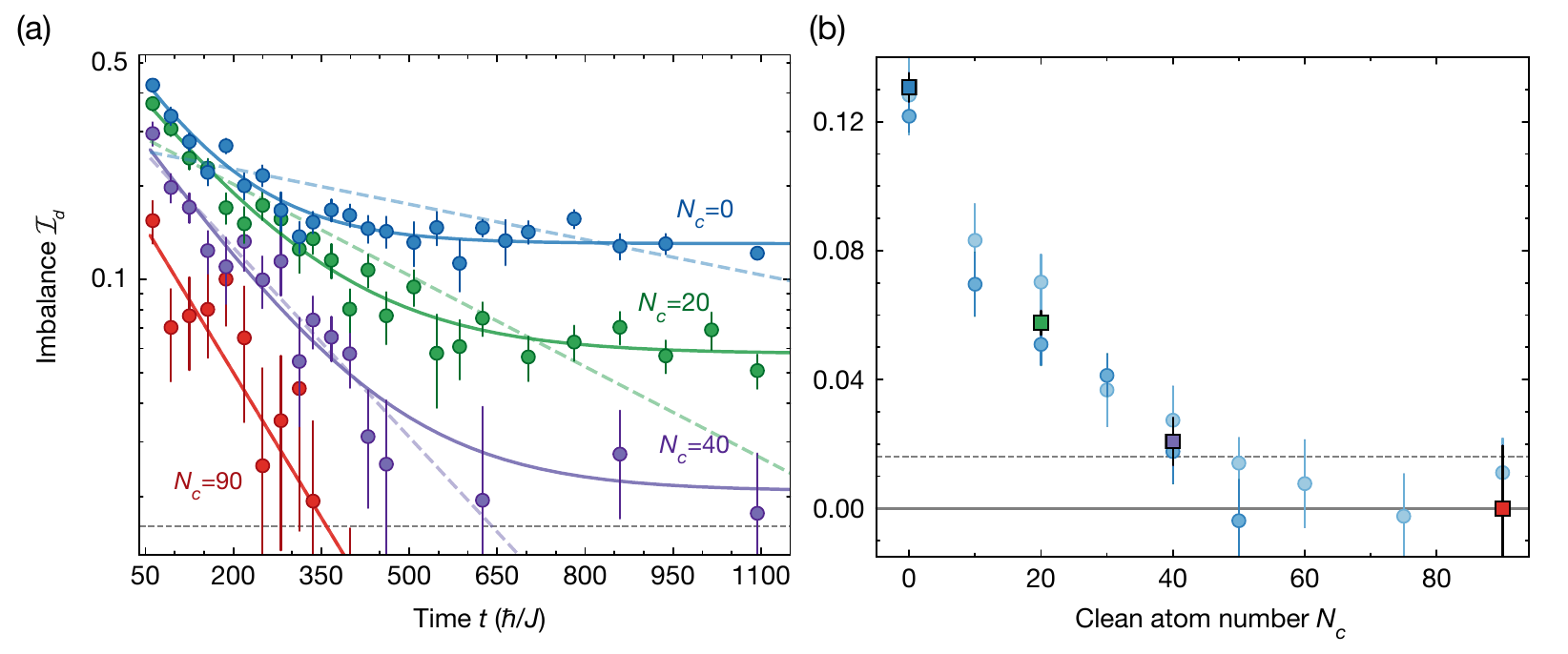}
\caption{ \label{fig:3}
\textbf{Dynamics and delocalization in the presence of a quantum bath.} (a) Dynamics of the dirty-state imbalance $\mathcal{I}_{d}$ for four
different bath sizes ($N_{c}=0$ in dark blue,
$N_{c}=20$ in green, $N_{c}=40$ in purple and $N_{c}=90$ in red). The dashed lines indicate exponential fits and the solid lines fits of an exponential with an offset. Introducing the clean component leads to delocalization, indicated by the reduced imbalance. The imbalance relaxes completely for the two largest bath sizes, while for the smallest size of the bath ($N_{c}=20$), a finite imbalance remains. (b) Experimental steady-state imbalance as a function of the bath size. The data was measured at $t=859\,\tau$ (round points in blue) and at $t=1094\,\tau$ (round points in light blue).  The square points correspond to the asymptotic offsets obtained from the four solid line fits in (a). The horizontal dashed gray line indicates the typical statistical threshold at which the imbalance is compatible with zero. The error bars indicate one standard deviation of the mean.}
\end{figure*}

We now turn to studying the effects of coupling a tunable atomic bath to the system. To this end, we initially prepare a mixture with a preset number $N_{c}$ of atoms in the clean state and after the dynamics, we ensure that the detection is only sensitive to atoms in state $\ket{d}$ by removing all $\ket{c}$ atoms with a resonant light pulse prior to detection (see Supplementary Information~\cite{SI}). The evolution of the imbalance $\mathcal{I}_d$ for three different bath sizes ($N_{c}=20,\, 40,\, 90$) is shown in Fig.~\ref{fig:3}(a), together with the dirty-only case as reference. Generally, the larger the bath, the smaller the imbalance in the long-time limit. While introducing a fraction of clean atoms also implies reducing the density of the dirty component, it is important to consider that this alone would actually yield a higher long-time imbalance~\cite{choi2016}. In the $N_{c}=90$ case, the imbalance relaxes to a vanishing value in less than $300\,\tau$, implying that the particles have delocalized over at least several lattice sites. A time constant of $140(30)\,\tau$ can be extracted from a single exponential fit (red dashed line) for this relaxation process. A similar delocalization takes place for the $N_{c}=40$ case, whose exponential fit gives a slower time constant of $200(20)\,\tau$. These results indicate that the clean component indeed acts as an effective bath, destabilizing and thermalizing the localized dirty component. Importantly, this effect is caused by collisional interactions only, whose strength is equal to the intracomponent interactions in the system, i.e. no new energy scale is introduced when adding the clean component. When reducing the size of the bath even further, to $N_{c}=20$, we still observe an overall reduction of the imbalance, but the dynamics is qualitatively different. A finite imbalance still persists for the longest measured times, and a simple exponential fit does not describe the dynamics accurately anymore. Introducing a steady-state offset, as in the case without a bath, matches the data significantly better (see supplementary information \cite{SI}), such that this is the simplest model describing all datasets (solid lines). A bootstrap analysis of the data after $625\,\tau$ allows us to bound the subsequent relaxation to be $t_{3}>1100\,\tau$ with a confidence of $92\%$.

The observed dynamics for smaller baths may be explained by an inefficient delocalization of the dirty component, with a decay rate much smaller than the other time scales in the system, but could also hint at a failure of thermalization. Theoretical studies for finite 1D systems with coupled clean and dirty components~\cite{nandkishore2015a, hyatt2017} found a persisting localization in certain regimes of reduced tunneling of the clean component, for which the coupling rate of spatially separated points is decreased. To quantify the delocalizing effect of the bath at long evolution time, we show the imbalance $\mathcal{I}_d$ as a function of the bath size (see Fig.~\ref{fig:3}(b)) for two different evolution times ($t=859\,\tau$ and $t=1094\,\tau$). The values at the two times are similar and they cannot be distinguished from the steady-state offsets obtained from the fits of Fig.~\ref{fig:3}(a). For bath sizes below $N_{c}\approx 40$ a finite imbalance $\mathcal{I}_d$ remains at long times.

\begin{figure}
\centering
\includegraphics{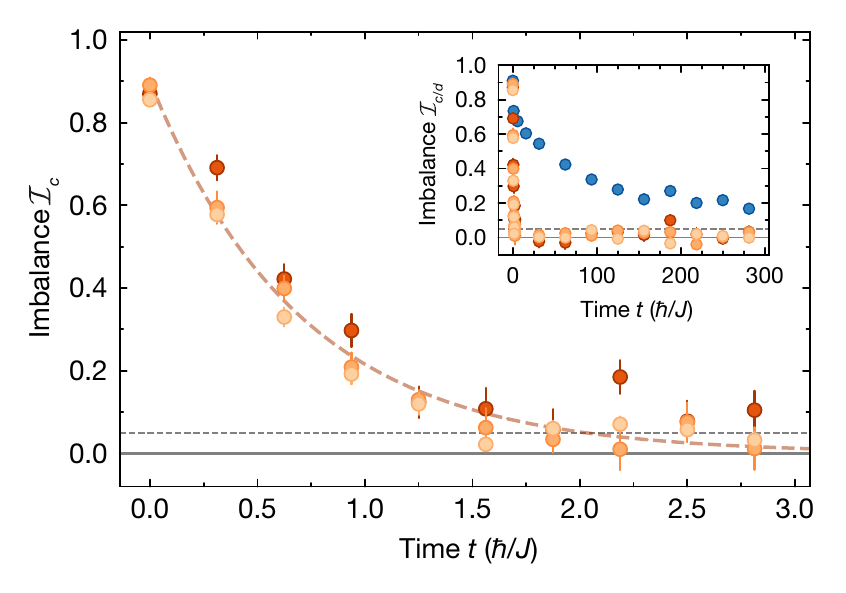}
\caption{ \label{fig:4}
\textbf{Imbalance dynamics of the clean component.} Evolution of
the clean component imbalance $\mathcal{I}_{c}$ for three different atom numbers
of the clean component ($N_{c}=120$ in light orange, $N_{c}=60$ in orange and
$N_{c}=20$ in dark orange). The dashed-line curve is an exponential fit with time constant of $0.7\,\tau$. All bath sizes result in a similar behavior, with the imbalance vanishing in few tunneling times. The inset shows
the long-time evolution of the clean component compared with the dynamics of
the dirty component in the $N_{c}=0$ case, highlighting the strong difference in the time scales. The horizontal dashed gray lines indicates
the typical statistical threshold at which the imbalance is compatible with
zero. The error bars represent one standard deviation of the mean.}
\end{figure}

Finally, to probe the back-action of the dirty component on the bath dynamics, we also tracked the evolution of the bath component in our
system. This was realized by removing the dirty component
$\ket{d}$ prior to detection, in order to detect the clean particles $\ket{c}$
only (Fig.~\ref{fig:4}). The results show that no matter how small the bath is, $\mathcal{I}_c$
relaxes very quickly, on a time scale of a few tunneling times $\tau$. Any potential interaction-induced localization of the small bath by the dirty component would therefore be characterized by a localization length spanning many sites, beyond what can be detected by the short-distance probing used here. Such a behavior is expected given that even the remaining imbalance of the dirty component $\mathcal{I}_d= 0.07(2)$ in the $N_{c}=20$ case is already very small, i.e. it can only serve as a rather weak disorder source.

The experiments reported in this work shed first light on MBL systems in contact with a quantum bath of tunable size. Generally, the presence of such a bath tends to drive the system towards delocalization, which eventually becomes complete even on our experimentally accessible time scales, when the bath formed by the clean component becomes large enough. Follow-up experiments may explore different initial states and disorder regimes
to settle the question of proximity-induced localization~\cite{nandkishore2015a, hyatt2017},
demonstrating the localization of a system due to and not despite of
interactions. Furthermore, the debated question on the stability of MBL in the
presence of thermal inclusions can be directly addressed in a similar
experiment with engineered low-disorder regions~\cite{agarwal2017, deroeck2017}.

\bigskip

\begin{acknowledgments}
\textbf{Acknowledgments:} We acknowledge fruitful discussions with D. A. Huse,
P. Bordia, H. L\"uschen, Y. Bar Lev and M. Heyl. Financial support was provided by the Max
Planck Society (MPG) and the European Union (UQUAM, Marie Curie Fellowship to J-y. C.).
\end{acknowledgments}

\renewcommand{\thefigure}{S\arabic{figure}}
\setcounter{figure}{0} 

\bigskip
\section*{Supplementary information}

\subsection*{Initial state preparation}

Our experiment began with the preparation of a unity-filling rubidium-87 Mott
insulator in a two-dimensional square optical lattice with a lattice spacing of $a_{\mathrm{lat}}=\SI{532}{nm}$~\cite{sherson2010}. The typically prepared system consists of 250 atoms, and we set the initial depth of the two lattices in the atomic plane to $40\,E_{\mathrm{r}}$, where $E_{\mathrm{r}}=h^{2}/8ma^{2}_{\mathrm{lat}}$ is the recoil energy, $h$ the Planck constant and $m$ the atomic mass. At
such a lattice depth, virtually no tunneling takes place, and thus the prepared state is separable.

We proceeded by removing all the atoms on the odd lattice sites along the $x$ axis, thereby preparing a charge-density wave (CDW) along one direction (see schematic in Fig.~1 of the main text). To do so, we use a spatially modulated laser beam with $\sigma^{-}$ polarization and a wavelength of \SI{787.55}{nm}, which induces a differential lightshift of $ h \times 10$ kHz between the two hyperfine states $\ket{c} = \ket{F=1, m_{F}=-1}$ and $\ket{d} = \ket{F=2, m_{F}=-2}$~\cite{weitenberg2011b}. We then apply a microwave sweep to transfer the illuminated atoms to the hyperfine state $\ket{d}$ and remove them by shining resonant light on the cycling transition of
the $\mathrm{D}_{2}$ line.  The initial state fidelity is characterized by the imbalance between the even $N_{e}$ and odd $N_{o}$ lattice sites $ \mathcal{I}
= (N_{\mathrm{e}}-N_{\mathrm{o}})/(N_{\mathrm{e}}+N_{\mathrm{o}})  = 0.91(1)$, and the remaining atom number is $N=124(12)$.

To prepare an admixture of the two hyperfine states $\ket{d}$ and $\ket{c}$, we afterwards apply a resonant microwave pulse of a certain length, which generates a state $\ket{ \mathrm{ \Psi }}_{\textit{\textbf{i}}} = \sqrt{1-\eta} \,
\ket{d}_{\textit{\textbf{i}}}+\sqrt{\eta}\,  \ket{c}_{\textit{\textbf{i}}} $ in each lattice site.
After less than one tunneling time, inhomogeneities due to the disorder
potential will lead to dephasing between the two spin states, and hence the
whole system can be treated as a statistical spin mixture with a fraction $\eta$ in the
$\ket{c}$ state.

\subsection*{Disorder potential}

After the preparation of a CDW, we quenched the system by ramping up a projected disorder potential and lowering the depth of the in-plane lattices from  $40\, E_{\mathrm{r}}$ to  $12\,E_{\mathrm{r}}$ in less than \SI{5}{ms}. The disorder potential is generated by the spatial modulation of a laser beam using a digital micromirror device (DMD), such that each lattice site of our 2D system features an individually programmable light shift~\cite{choi2016}. The DMD consists of a $1024 \times 768$ micromirror array with a $\SI{13.7}{\micro \meter}$ micromirror pitch, and approximately $7 \times 7$ of these mirrors oversample the point spread function (Gaussian with $\sigma=0.48(1)\, a_{\mathrm{lat}}$) of our system. To image the disorder onto the atoms we use a high-resolution objective with a numerical aperture of $\mathrm{NA}=0.68$~\cite{sherson2010}. A pseudorandom number generator is used to produce a 2D random pattern, which is chosen different for every experimental realization. We resolved the microwave resonances spatially, which is equivalent to extracting the local light shift~\cite{choi2016}. The histogram of all the local shifts displays a Gaussian distribution with full width at half maximum $\Delta$, which characterizes the strength of the disorder. Our imaging system introduces a finite correlation to the disordered potential due to its finite resolution, for which we measure a correlation length of $0.63(1)\,a_{\mathrm{lat}}$~\cite{choi2016}. The disorder potential can locally modify the Bose-Hubbard tunneling parameter $J$, in contrast to the bare lattice case, but in our parameter regime we expect this effect to be reasonably small. The disorder strength is much smaller than the lattice depth ($\Delta\approx 0.02 \, V_{\mathrm{lat}}$), and the modification of the tunneling $\delta J$ will be well below $0.08J$. 

The disorder beam is tuned to the so-called `tune-out' wavelength of the $\ket{F=1, m_{F}=-1}$ state, such that species-dependent potentials can be tailored~\cite{Leblanc2007}. This allows for tuning to a configuration where the $\ket{c}$ component experiences a vanishing light shift, while the light shift for the $\ket{d}$ state leads to an attractive potential. Aside from the programmed on-site disorder potential $\delta_{i}$, all atoms are equally sensitive to an overall harmonic trapping potential $V_{\textit{\textbf{i}}} = m
a^{2}_{\mathrm{lat}} (\omega^{2}_{x} x^{2}_{\textit{\textbf{i}}} + \omega^{2}_{y} y^{2}_{\textit{\textbf{i}}})/2 $
with frequencies $(\omega_{x},\omega_{y}) = 2 \pi \times (51(2),55(2))$Hz.

\subsection*{Measuring the occupation number}

At the end of each run we measure the atomic occupation in each lattice site. We first freeze the tunneling motion by increasing the lattice depth to $60\,E_{\mathrm{r}}$ in less than half a millisecond. To selectively measure only one of the two hyperfine components, we then remove all the atoms in the state that we do not wish to measure. To image the $\ket{c}$ state, we push the $\ket{d}$-state atoms using a resonant $\mathrm{D}_{2}$ light pulse, while for detecting the $\ket{d}$ state, we apply a microwave sweep to swap the two hyperfine spin states before the optical push-out pulse, thus removing the atoms which were originally in the $\ket{c}$ state. 

After the state selection, all lattices are ramped to their maximum depth and an optical molasses is used to scatter fluorescence photons and simultaneously cool the atoms. We expose an EMCCD camera for \SI{1}{s}, thereby obtaining single-site-resolved images of the parity-projected density distribution~\cite{sherson2010}.

\begin{figure}
\centering
\includegraphics{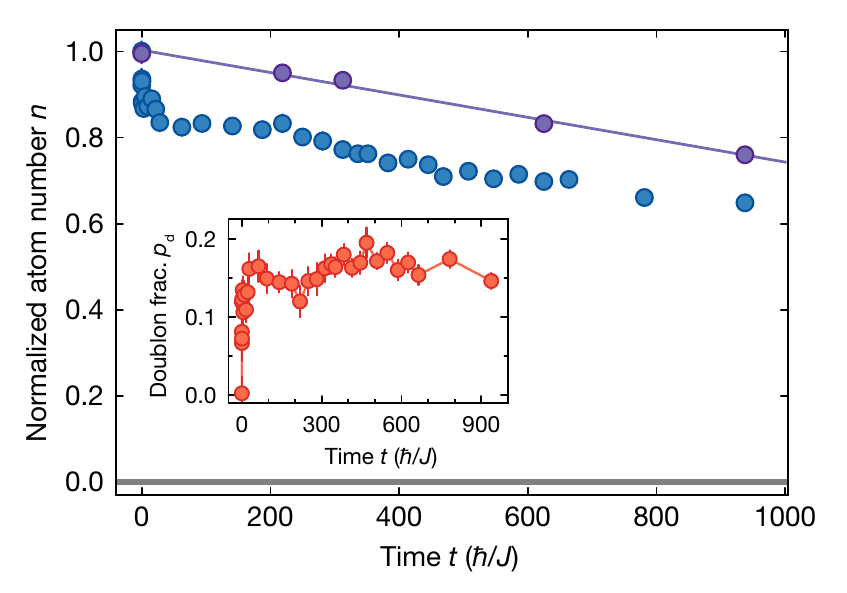}
\caption{ \label{fig:S1}
\textbf{Atom loss and doublon formation.}
The purple points show the normalized total atom number $n$ as a function of holding time. The purple line indicates a linear fit to the decay, with a typical time of $3300(300)\,\tau$. In blue we show the normalized parity-projected atom number $n_{\mathrm{CDW}}$ for the CDW evolution as a function of evolution time. The inset shows in red the estimated doublon fraction $p_{\mathrm{d}}$, obtained from the subtraction of the plotted normalized atom numbers.}
\end{figure}

\subsection*{Atom loss and doublon generation}

An estimate of the total atom number in the system cannot be directly obtained from the in-situ fluorescence imaging, since parity projection prevents us from distinguishing empty lattice sites from doubly-occupied ones. When discarding local information, however, we can circumvent this issue. Performing a short time of flight in the 2D plane right before imaging dilutes the atom density, such that essentially no doubly-occupied sites remain in the final atomic distribution.  We used this technique to measure the  atom number of an initially prepared Mott insulator by turning both in-plane lattices off for \SI{1}{ms}. By measuring the total atom number $N(t)$ for different holding times, we were able to estimate the atom losses. This is shown in Fig. S1, in which the purple points represent the normalized atom number, $n(t)=N(t)/N(0)$, and the purple line is a linear fit with an atom number decay of $3300(300)\,\tau$. This means that after $ 600 \,\tau$, approximately $17\,\% $ of the initially prepared atoms have been lost, for which the main loss mechanism is induced by technical fluctuations of the lattice-beam intensities~\cite{choi2016}. We estimate the photon scattering rate from the disordered potential to be below $\gamma = 7 \cdot 10^{-5}\,\tau^{-1} $. Though our atom loss is comparable to previous work in other quantum-gas experiments~\cite{lueschen2017a}, the measured imbalance decay seems little affected by it.

By comparing the parity-projection-free atom number with the one obtained from direct in-situ measurements of the CDW relaxation (Fig. 2 in the main text), we can additionally estimate the fraction of dynamically generated doublons even for long times. We define the doublon fraction as $p_{\mathrm{d}}(t)=2 \cdot N_{\mathrm{dou}}(t)/N(t)=(N(t)-N_{\mathrm{CDW}}(t))/N(t)=(n(t)-n_{\mathrm{CDW}}(t))/n(t)$, where $ N_{\mathrm{dou}}(t)$ is the number of doublons, and $N_{\mathrm{CDW}}(t)$ and $n_{\mathrm{CDW}}(t)$ are respectively the absolute and normalized parity-projected atom number. For the data measured at $\Delta=28\,J$, we find that the doublon fraction remains approximately constant after the fast initial formation (see Fig. S1). We did not take into account the effect of triply-occupied sites, which at the current experimental settings is estimated to be negligible.

\begin{figure*}
\centering
\includegraphics{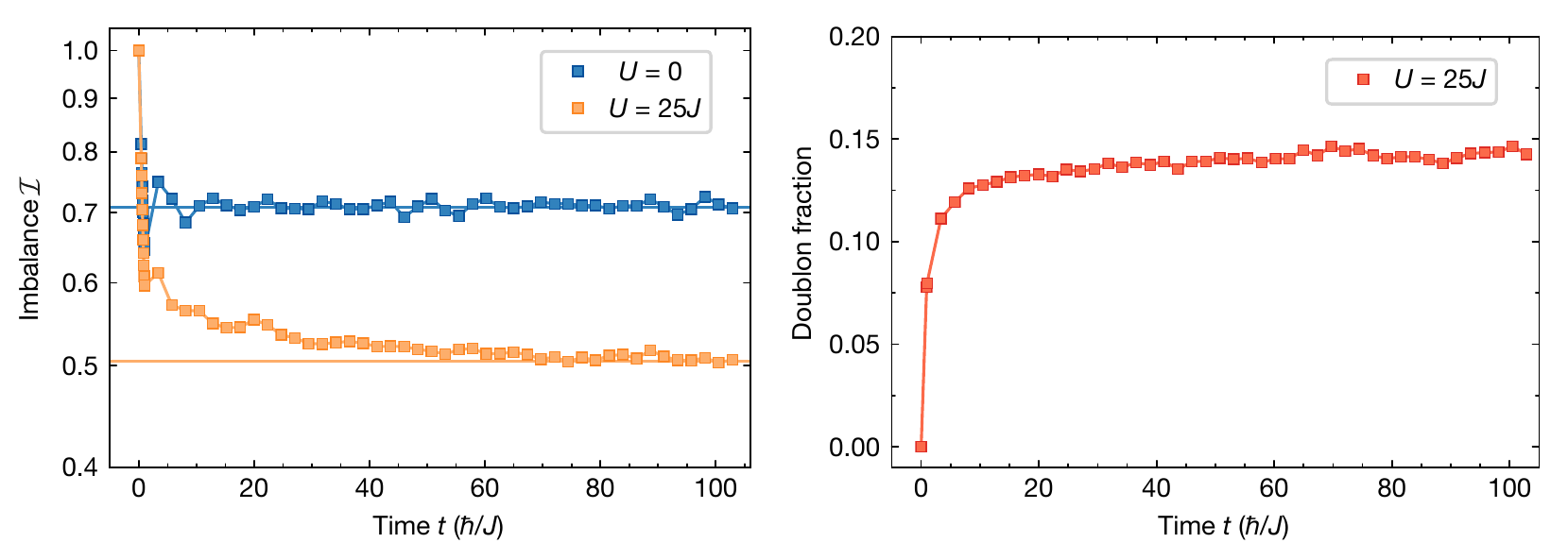}
\caption{ \label{fig:S2}
\textbf{Results of the dynamics obtained by exact diagonalization.}
In the left we plot the evolution of the imbalance for $\Delta=25\,J$ for a non-interacting ($U=0$) and an interacting ($U=25\,J$) case, in which we can appreciate the slow relaxation when interactions are present. The second figure shows the evolution of the doublon fraction, which essentially saturates after a very fast initial formation.}
\end{figure*}

\subsection*{Numerical simulations}

To get more insight on the slow relaxation observed for the dirty-component imbalance dynamics (Fig. 2 in  the main text) we have performed numerical simulations based on exact diagonalization for a small disordered-Bose-Hubbard system, for which we used the Quspin package~\cite{Quspin}. The considered system is a ladder of $2 \times 6$ lattice sites with periodic boundary conditions, populated with 5 particles in a CDW-like pattern. We choose the system parameters close to the experimental ones, with a disorder distribution given by a Gaussian with full width at half maximum of $\Delta=25\,J$ and we have considered both a non-interacting ($U = 0$) and an interacting case ($U = 25\,J$). 

Without interactions, the imbalance dynamics shows a very fast relaxation to $\mathcal{I} = 0.7$, which takes less than $10\,\tau$ (see Fig. S2). In contrast, for the interacting system a steady state is only reached after almost $100\,\tau$, and one can distinguish a very fast initial decay to $\mathcal{I} = 0.6$ in few tunneling times from a much slower relaxation afterwards, quite similar to our experimental findings. Interestingly, such a slow relaxation is not observed in one-dimensional numerical simulations with the same interaction strength, not even for an intermediate disorder strength $\Delta$. The slow decay stops at a steady state imbalance of $\mathcal{I} \approx 0.5$, which is notably higher than the one measured in our experiment ($\mathcal{I} \approx 0.15$). Such a discrepancy could be explained by finite-size effects due to the small number of sites and atoms in the simulation in comparison to the real system. From our simulations, we have also obtained the doublon-fraction dynamics, which clearly indicate a fast doublon formation during the initial time scale, followed by a much slower increase, also in a similar fashion to the experimental results.

\begin{figure}[b]
\centering
\includegraphics{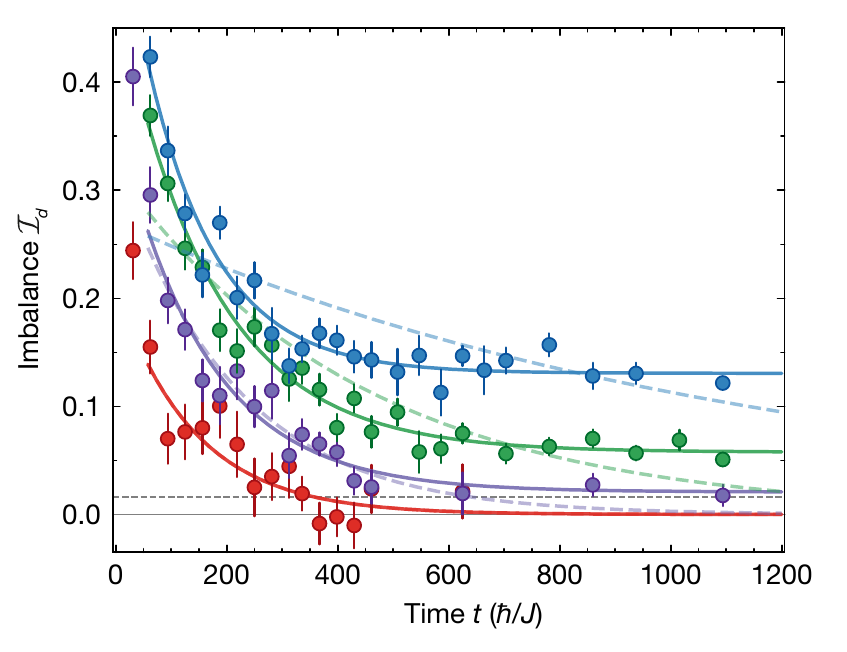}
\caption{ \label{fig:S3}\textbf{Delocalization dynamics in a lin-lin plot.}  Shown
is the dynamics of the dirty-state imbalance $\mathcal{I}_{d}$ for four
different bath sizes ($N_{c}=0$ in blue,
$N_{c}=20$ in dark green, $N_{c}=40$ in purple and $N_{c}=90$ in red). The dashed lines indicate exponential fits and the solid lines fits of an exponential with an offset. The horizontal dashed gray line indicates the typical statistical
threshold at which the imbalance is compatible with zero. The error bars
indicate one standard deviation of the mean.}
\end{figure}

\subsection*{Goodness of fit estimation}

Here we provide a quantitative estimation of the goodness of the fits in Fig. 3 of the main text. We have calculated the reduced chi-square statistic $\chi^{2}_{\nu}$ for the single-exponential ($\mathcal{I}(t)=A_{1} e^{-t/t_{1}}$) and exponential-with-an-offset ($\mathcal{I}(t)=A_{1} e^{-t/t_{1}}+A_{2} $) fits for the four datasets (see Tab. 1). The values show that while the single exponential fit has a good agreement with the data for the two cases of bigger bath size, this gets worse as the bath size is reduced, as seen by the higher $\chi^{2}_{\nu}$ values for $N_{c}= 20$ and the bath-free case $N_{c}= 0$. Instead, the fit of an exponential with an offset provides a good fit for all four cases. We also show Fig. 3 with linear axes in Fig. S3, including points (for the largest bath case) which are not included in the log-lin plot of the main text due to diverging errorbars.

\begin{table}[h]
\begin{tabular}{lllll}
\hline
\multicolumn{1}{|l|}{Clean atom number $N_{c}$}              & \multicolumn{1}{c|}{90} & \multicolumn{1}{c|}{40} & \multicolumn{1}{c|}{20} & \multicolumn{1}{c|}{0} \\ \hline
\multicolumn{1}{|l|}{Single exponential $\chi^{2}_{\nu}$}          & \multicolumn{1}{l|}{0.88}  & \multicolumn{1}{l|}{1.14}  & \multicolumn{1}{l|}{5.12}  & \multicolumn{1}{l|}{9.05} \\ \hline
\multicolumn{1}{|l|}{Exponential with an offset $\chi^{2}_{\nu}$} & \multicolumn{1}{l|}{0.96}  & \multicolumn{1}{l|}{0.99}  & \multicolumn{1}{l|}{0.99}  & \multicolumn{1}{l|}{1.6} \\ \hline
                                    &                         &                         &                         &                       
\end{tabular}
\caption{\textbf{Reduced chi-squared statistics $\chi^{2}_{\nu}$ for the two different fits of Fig. 3 and Fig. S3
}}
\end{table}

\bigskip

%\textbf{Author Contributions:} All authors contributed extensively to the
%planning, data acquisition and analysis of the results presented here.

\bibliography{bibliography}

\end{document}